\def\bra#1{\left<#1\right|}
\def\ket#1{\left|#1\right>}
\def\mydiv{\mathop{\rm div}\nolimits}%

\documentstyle[twocolumn,eqsecnum,aps,prl,epsfig,floats]{revtex}  
\begin{document}
\draft
\title{ \hfill {\small HLRZ1999\_8, PITHA 99/7}\\
  Scaling analysis of the magnetic monopole mass and condensate\\
  in the pure U(1) lattice gauge theory
}
\author{J.~Jers{\'a}k, T.~Neuhaus, and H.~Pfeiffer}
\address{Institut f{\"u}r
Theoretische Physik E, RWTH Aachen, Germany}
%
\date{\today}
\maketitle
\begin{abstract}
  We observe the power law scaling behavior of the monopole mass and
  condensate in the pure compact U(1) gauge theory with the Villain action. In
  the Coulomb phase the monopole mass scales with the exponent $\nu_{\rm
    m}=0.49(4)$.  In the confinement phase the behavior of the monopole
  condensate is described with remarkable accuracy by the exponent $\beta_{\rm
    exp}=0.197(3)$.  Possible implications of these phenomena for a
  construction of a strongly coupled continuum U(1) gauge theory are
  discussed.

\end{abstract}

\section{Introduction}

The phase transition between the confinement and Coulomb phases of the
strongly coupled pure compact U(1) lattice gauge theory (compact QED) has
recently received renewed interest and two of its aspects were investigated in
large numerical simulations. First, several attempts have been made to
distinguish between the weak first order and second order scenarios for the
Wilson action and in the extended coupling parameter space. The question is
whether the two-state signal decreasing slowly with increasing lattice volume
extrapolates to a nonzero or zero value, respectively, in the thermodynamic
limit. (For a recent discussion of this subject and earlier references see Ref.
\cite{ArLi99}).

Second, a scaling behavior of various bulk quantities and of the gauge-ball
spectrum consistent with a second order phase transition and universality has
been observed in the vicinity of some points on the manifold separating the
confinement and Coulomb phases, outside their narrow neighbourhood in which
the two-state signal occurs \cite{JeLa,CoFr97b,CoJe99a,CoJe99b}. This suggests
that there may exist regions of the parameter space where the transition is
continuous, though it may be not such for the particular action used in the
simulation.

A continuous transition would allow the construction of a continuum theory.
But even if no critical point exists, the theory might be considered as an
effective theory, with finite but large cutoff, provided the range of scales
at which a second order-like behavior holds is large. The question is then
whether such a (possibly effective) continuum U(1) theory would be interesting
in some sense, e.g. would it have a phase transition, confinement, etc., in
analogy to the lattice regularized theory.

In this paper we address this second aspect and extend the investigation of
the scaling behavior to observables related to the magnetic monopoles. To our
knowledge this subject has not yet been investigated in a systematic way. But
the issue is important, as whatever is interesting in the compact lattice QED
is essentially related to the monopoles: The phase transition itself is known
to be associated with the occurence of magnetic monopoles being topological
excitations of the theory \cite{Po,Pe78,PoPo91}. Modifications of the monopole
contribution to the action have appreciable consequences for its position
\cite{BaShr} and properties \cite{BoLi}. The long distance force in the
confinement phase \cite{StTh,DiLu98} and the chiral symmetry breaking
\cite{BiHa98} are best understood in terms of the monopole condensate. The
charge renormalization in the Coulomb phase is due to the antiscreening by
monopoles \cite{JeNe85,Lu90,PoWi91}. Thus the existence of an interesting
effective U(1) theory presumably depends on whether the monopoles persist to
play an important role in it, i.e.  on the scaling behavior of the monopoles.

Our findings are as follows: In the Coulomb phase we find at various values of
the coupling $\beta$ a very clean exponential decay of the monopole
correlation function in a large range of distances. This demonstrates the
dominance of a single particle state in this correlation function, the
monopole, whose mass we determine. The monopole mass extrapolated to the
infinite volume, $m_\infty$, scales with the distance from the phase
transition as
\begin{equation}
       m_\infty(\beta)=a_m{(\beta-\beta_c)}^{\nu_{\rm m}},
       \label{m_infty}
\end{equation}
where 
\begin{equation}
       \nu_{\rm m}=0.49(4).
       \label{nu_m}
\end{equation}

This value of the exponent $\nu_{\rm m}$ can be compared with the values for
the correlation length exponents $\nu$ obtained for other observables. One of
these values is the non-Gaussian value
\begin{equation}
     \nu_{\rm {ng}} \simeq 0.35
     \label{nu_ng}
\end{equation}
found for the Lee-Yang zeros \cite{JeLa} at the transition, several gauge
balls \cite{CoFr97b,CoJe99a} and approximately also for the string tension
\cite{CoFr97b,CoJe99b}. The other, Gaussian value is
\begin{equation}
    \nu_{\rm {g}} \simeq 0.5.
     \label{nu_g}
\end{equation}
It has been observed previously for the scalar gauge ball
\cite{CoFr97b,CoJe99a} in the confinement phase.

Our main result is that the value (\ref{nu_m}) of $\nu_{\rm m}$ is
significantly greater than (\ref{nu_ng}) and consistent with (\ref{nu_g}).
This implies that monopoles would stay important in any of the conceivable
scenarios for a construction of an effective continuum U(1) theory.  If such a
theory were to be constructed in such a way that masses and other dimensionful
observables scaling with the non-Gaussian exponent $\nu_{\rm {ng}}$ were kept
finite nonzero in physical units, then the monopoles in the Coulomb phase
would get massless. Even if instead the Gaussian exponent $\nu_{\rm {g}}$ were
used, the monopole mass can be fixed at a finite value in physical units.
This second possibility might be particularly suitable in the Coulomb phase,
where no other scales are known.

In the confinement phase we have determined the scaling behavior of
the monopole condensate extrapolated to the infinite volume, $\rho_\infty$, to
be
\begin{equation}
\label{condscale}
  \rho_\infty=a_\rho{(\beta_c-\beta)}^{\beta_{\rm exp}},
\end{equation}
where
\begin{equation}
\label{magneticexp}
  \beta_{\rm exp}=0.197(3).
\end{equation}
The function (\ref{condscale}) describes extremely well the data in a broad  
interval and the scaling behavior of the condensate is thus well
established. 

However, the value of the magnetic exponent $\beta_{\rm {exp}}$ alone is not
sufficient for considering the continuum limit. For this purpose a
renormalised condensate is needed. A natural procedure (like e.g.~in the
broken $\phi^4$ theory) would be to find a pole in the monopole correlation
function in the confinement phase. We find a contribution suggesting such a
pole, but the data are consistent with its amplitude, as a function of the
lattice volume, extrapolating to zero in the thermodynamic limit. Thus
currently the results (\ref{condscale}) and (\ref{magneticexp}) do not allow a
conclusion about the condensate in a would-be continuum limit. They constitute
only a necessary step in this direction.

The results are presented as follows: In Sec. 2 we summarize known facts about
the ${\mathbf Z}$ gauge theory, which we use in the simulation. It is a dual
equivalent to the U(1) lattice gauge theory with the Villain action which we
actually investigate. In the Z gauge theory the monopole correlation functions
have a form originally found by Fr\"ohlich and Marchetti \cite{FrMa},
which is convenient for measurements. Antiperiodic boundary conditions
\cite{Wi93,PoWi91} are used allowing the consideration of a single monopole in
a finite volume in agreement with the Gauss law. These b.c. reduce the
magnetic U(1) symmetry \cite{FrMa,PoWi91} to Z$_2$. It is pointed out
that this symmetry is broken in the confinement phase, which necessitates some
caution during simulations.

In Sec. 3 we present our results for the monopole mass in the Coulomb
phase and determine its scaling behavior. In Sec. 4 the necessary
extrapolation procedure of the monopole mass results to the infinite
volume limit is discussed. The finite size effects are sizeable, as
monopoles with their Coulomb field are extended objects. The leading
term in the volume dependence can be determined from electrostatic
considerations, however, and further terms obey a simple ansatz.

In Sec. 5 we present the results for the monopole condensate in the
confinement phase and describe the search for a monopole condensate
excitation. We discuss our results and conclude in Sec. 6. 

\section{The dual formulation of pure U(1) lattice gauge theory}

\subsection{${\mathbf Z}$ gauge theory}
The partition function of pure U(1) lattice gauge theory
\begin{equation}
     \label{partition}
       Z=\int{\prod_{x\mu}}\int_{-\pi}^\pi d\theta_{x\mu}\,
       \exp\Bigl(-\sum_Ps(\theta_P)\Bigr)
\end{equation}
with an action $s(\theta_P)$, $\theta_P$ denoting the plaquette angles, is
related to the ${\mathbf Z}$ (integer) gauge theory by an exact duality
transformation \cite{Pe78,FrMa}. In the ${\mathbf Z}$ gauge theory, the link
variables $n_{x\mu}$ have integer values, and the partition function reads
\begin{eqnarray}
   \label{dual_partition}
        Z&=& \int{\mathcal D}\,n\,\exp\Bigl(-\sum_Ps^\ast(n_P)\Bigr), \\
   \label{measure}
      \int{\mathcal D}\,n &=& \prod_{x\mu}\sum_{n_{x\mu}=-\infty}^\infty,
\end{eqnarray}
with 
\begin{equation}
      n_P=n_{x\mu}+n_{(x+\hat\mu)\nu}-n_{(x+\hat\nu)\mu}-n_{x\nu}
\end{equation}
being the plaquette integer number associated with a plaquette at position $x$
and orientation $(\mu,\nu)$. In~(\ref{dual_partition}), $s^\ast(n_P)$ denotes
the dual action. The dual theory is a gauge theory invariant under the
transformations
\begin{equation}
   n_{x\mu}\mapsto n_{x\mu}+{(\nabla_\mu\ell)}_x=n_{x\mu}+\ell_{x+\hat\mu}-\ell_x
\end{equation}
with an integer valued function $\ell_x$ of the lattice points $x$.  The
theories (\ref{partition}) and (\ref{dual_partition}) are strictly equivalent
in the infinite volume limit and they should be comparable in large volumina.

The dual action associated with the Villain action
\begin{equation}
  \label{villain}
 s(\theta_P)=-\log\sum_{k=-\infty}^\infty\exp
 \Bigl(-\frac{\beta}{2}{(\theta_P+2\pi\,k)}^2\Bigr)
\end{equation}
is
\begin{equation}
  \label{dual_villain}
        s^\ast(n_P)=\frac{1}{2\beta}n_P^2
\end{equation}
whereas the extended Wilson action
\begin{equation}
       s(\theta_P)=\beta\,\cos(\theta_P)+\gamma\,\cos(2\theta_P)
\end{equation}
corresponds to the dual action
\begin{equation}
  \label{dual_exWilson}
      s^\ast(n_P)=-\log\Bigl(\int_{-\pi}^\pi
dz\,\cos(z\,n_P)e^{\beta\,\cos(z)+\gamma\,\cos(2z)}\Bigr). 
\end{equation}
Obviously, for a numerical simulation of the ${\mathbf Z}$ gauge theory the
dual Villain action (\ref{dual_villain}) is much more practical than
(\ref{dual_exWilson}). This is the reason we choose the Villain action
(\ref{villain}) in our current work.

\subsection{Dual correlation functions}
\label{dual_corr}

The magnetic monopoles in the U(1) lattice gauge theory are described by
fields $\Phi(x)$ whose correlation functions are defined by certain
modifications of the partition function (\ref{dual_partition})~\cite{FrMa}
\begin{eqnarray}
\label{disorder}
 \left<\Phi^\star(y_1)\cdots\Phi^\star(y_j)\Phi(z_1)\cdots\Phi(z_\ell)\right>\cr
  =\frac{1}{Z}\int{\mathcal D}\theta\,\exp\Bigl(-\sum_P s(\theta_P+X_P)\Bigr).
\end{eqnarray}
Here the magnetic (in the U(1) language) flux $X_P$ is generated by the
magnetic current density
\begin{eqnarray}
\label{currdens}
  \tilde J_{x\mu\nu\lambda}=X_{(x+\hat\lambda)\mu\nu}-X_{x\mu\nu}
  +X_{(x+\hat\mu)\nu\lambda}\cr
  -X_{x\nu\lambda}+X_{(x+\hat\nu)\lambda\mu}-X_{x\lambda\mu}
\end{eqnarray}
with sources $2\pi$ at $z_1,\ldots,z_\ell$ and $-2\pi$ at $y_1,\ldots,y_j$.
For given sources, the existence of $X_P$ on finite lattices depends on the
topology of the lattice, i.e.\ on the boundary conditions (b.c.).

In the dual formulation, the expression~(\ref{disorder}) becomes a usual
expectation value 
\begin{eqnarray}
\label{disorder_dual}
  \left<\Phi^\star(y_1)\cdots\Phi^\star(y_j)\Phi(z_1)\cdots\Phi(z_\ell)\right>\cr
  =\int{\mathcal D}n\,\exp\Bigl(-i\sum_{x\mu}J_{x\mu}n_{x\mu}\Bigr)\,
  \exp\Bigl(-\sum_Ps^\star(n_P)\Bigr)
\end{eqnarray}
of a non-local observable~\cite{FrMa}
\begin{equation}
\label{charge_op}
  \exp\Bigl(-i\sum_{x\mu} J_{x\mu}n_{x\mu}\Bigr),
\end{equation}
in which $J_{x\mu}$ is given by the current density~(\ref{currdens})
\begin{equation}
  J_{x\rho}=\frac{1}{6}\tilde J_{x\mu\nu\lambda}\epsilon_{\mu\nu\lambda\rho}.
\end{equation}
The observable in~(\ref{disorder_dual}) resembles the non-local expression of
a charge operator in terms if its Coulomb field. In the pure U(1) gauge theory
an analogous expression would not be gauge invariant. However, in the case of
${\mathbf Z}$ gauge theory, (\ref{charge_op}) is gauge invariant because the
sources of $J_{x\mu}$ are integer multiples of $2\pi$ \cite{FrMa}.

\subsection{The monopole observable}

To determine the correlation functions~(\ref{disorder_dual}) in a Monte-Carlo
simulation, it is useful to choose the field $X_P$ so that $J_{x\mu}$ vanishes
on all links in a certain direction $t$ (referred to as the ``time'' direction)
\cite{PoWi91}. Then the observable~(\ref{charge_op}) decomposes into factors
that are local in $t$ and nonlocal in the ``space'' directions $\vec x$. 

Therefore one defines $J_{x\mu}$ for each source located at $(\vec x_0,t_0)$
as 
\begin{equation}
 \label{sources_given}
   J_0(\vec x,t)=0,\qquad \vec J(\vec x,t)=\vec B(\vec x)\,\delta_{tt_0}
\end{equation}
where $\vec B(\vec x)$ is a solution of
\begin{equation}
 \label{div_equation}
   \mydiv\vec B(\vec x)=2\pi\,\delta_{\vec x\vec x_0}
\end{equation}
in three dimensions. Such a $J_{x\mu}$ in~(\ref{sources_given}) yields the
monopole field
\begin{equation}
 \label{monopole}
  \Phi(\vec x_0,t_0)=\exp\Bigl(-i\sum_{x\mu}J_{x\mu}n_{x\mu}\Bigr)
\end{equation}
Products of these fields with sources at different positions therefore lead to
superpositions of currents $J_{x\mu}$ in~(\ref{disorder_dual}) and
(\ref{charge_op}). 

In the case of a source (\ref{sources_given}) and (\ref{div_equation}) the
boundary conditions in the three space dimensions have to be chosen
appropriately.  Whereas there is no solution of~(\ref{div_equation}) with
periodic boundaries, a solution exists for antiperiodic boundaries
\cite{PoWi91}
\begin{equation}
\label{boundary_def}
  n_{\mu}(\vec x+L\vec e_j,t)=-n_{\mu}(\vec x,t)
\end{equation}
on the lattice of spacial extension $L$. That solution can be obtained by the three-dimensional discrete
Fourier-transformation of~(\ref{div_equation}) on the lattice with respect to
$L$-antiperiodic base functions.  In terms of the monopole
observable~(\ref{monopole}), the boundary conditions~(\ref{boundary_def})
correspond to magnetic charge conjugation $C$~\cite{PoWi91} and are therefore
called $C$-periodic.

In principle it would be possible also for periodic b.c. to correlate
monopole-antimonopole pairs at large relative distance and determine
approximately twice their mass. However, antiperiodic b.c. are much more
practical allowing to determine the mass of a single monopole on lattices of
moderate sizes.

\subsection{Dispersion relation and finite size effects}

The real and imaginary part of the monopole are even resp.\ odd with respect
to $C$. The observables with definite momentum are $C$-even 
\begin{equation}
\label{even_part}
  \Phi_+(\vec p,t_0)=\sum_{\vec x_0}\Re(\Phi(\vec x_0,t_0))e^{i\vec p\cdot\vec
  x_0}
\end{equation}
and $C$-odd
\begin{equation}
\label{odd_part}
  \Phi_-(\vec p,t_0)=\sum_{\vec x_0}\Im(\Phi(\vec x_0,t_0))e^{i\vec p\cdot\vec
  x_0}.
\end{equation}

Since Fourier-transformation in~(\ref{even_part}) is done with $L$-periodic
functions, the momenta are $\vec p=(p_1,p_2,p_3)$ with $p_j=2\pi k_j/L$ and
integer $k_j$. In~(\ref{odd_part}) Fourier-transformation is applied to
$L$-antiperiodic functions, so that the momenta are $p_j=(2k_j+1)\pi/L$ with
integer $k_j$. As a consequence, the whole monopole observable
$\Phi=\Phi_++\Phi_-$ has no well-defined dispersion relation on finite volume
lattices. Only in the infinite volume limit, the momentum spectra degenerate
and the monopole mass determined by means the even and odd observables
(\ref{even_part}), (\ref{odd_part}) lead to the same mass. We will have to
take this fact into consideration when we examine the volume dependence of our
results in Sec.~\ref{volume}.

The monopole masses on finite lattices are determined using 
\begin{equation}
  \label{m_+}
m_+=E_1^+ 
\end{equation}
in the correlation function of the operator (\ref{even_part}). In the case of
(\ref{odd_part}), we assume that the mass is obtained from the energy by the
usual particle dispersion relation
\begin{equation}
\label{dispersion}
  E_1^2=m_-^2+3\,{(2\sin\frac{\pi}{2L})}^2.
\end{equation}
However, for a particle with a Coulomb field in a finite volume this relation
is presumably only an approximation. We take the possible corrections into
account in a phenomenological extrapolation to the infinite volume.  Such a
way of extrapolation to the infinite volume is necessary anyhow, as the finite
size effects for a particle with extended Coulomb field are large and only
partly under analytic control. The results do not change significantly if we
use the lattice dispersion relation with $2(\cosh(E)-1)$ instead of $E^2$.

The magnetic U(1) symmetry is reduced to a ${\mathbf Z}_2$ symmetry by
imposing $C$-periodic boundary conditions \cite{PoWi91,PoPo91}. In the
confinement phase of pure U(1) gauge theory, this remaining ${\mathbf Z}_2$
symmetry is spontaneously broken.  This ${\mathbf Z}_2$ symmetry corresponds
to the sign of the $\left<\Phi_+\right>$ expectation value whereas
$\left<\Phi_-\right>=0$ due to $C$-antisymmetry. On finite lattices flips
between both signs of $\left<\Phi_+\right>$ can occur, distorting the
measurements. Therefore we discard parts of the runs where such flips occur.

\section{Results in the Coulomb-phase}

\subsection{Simulations and statistics}
\label{simstat}

We simulate the ${\mathbf Z}$ gauge theory with the
action~(\ref{dual_villain}). The boundary conditions are $C$-periodic in
spacial directions and periodic in time direction. We use the same heat bath
update as the authors of~\cite{PoWi91} together with a new implementation of
the monopole observable~(\ref{monopole}) optimized for vectorization on a
Cray~T90.

The lattice volumes are $L^3\cdot T$ with $T=28$ fixed and different
$L$ ranging from $4$ to $18$. Both monopole operators~(\ref{even_part})
and~(\ref{odd_part}) are measured after each $25$ update sweeps for each
time slice. From these data, the correlation functions are computed. In
addition, we determine the action density. 

\begin{table}[ht]
\begin{center}
\begin{tabular}{lrr}
$\beta$ & $L_{\rm max}$ & $\tau_{\rm int}$ \\
\hline
0.645   & 18 & 11.0 \\
0.647   & 18 &  2.5 \\
0.65    & 18 &  2.3 \\
0.654   & 18 &  1.9 \\
0.66    & 16 &  1.3 \\
0.668   & 16 &  1.1 \\
0.678   & 16 &  1.0 \\
\end{tabular}
\end{center}
\caption{
  \label{autocorr} Values of $\beta$ at which the simulations in the Coulomb
  phase were performed. The lattice sizes are $L=4,6,\ldots,L_{\rm max}$. At
  each $\beta$, at least $10^4$ measurements have been made. $\tau_{\rm int}$
  is the integrated autocorrelation time of the action density in multiples of
  $25$ sweeps determined on the $L=8$ lattice.}
\end{table}

Table~\ref{autocorr} gives the simulation points in the Coulomb phase, the
range $L=4,6,\ldots,L_{\rm max}$ of lattice sizes, and the integrated
autocorrelation times of the action on the $L=8$ lattice. To thermalize the
system, we skipped the first $2.5\cdot 10^3$ (far away from the phase
transition) to $2.5\cdot 10^4$ sweeps (close to the phase transition). At each
value of the parameters, at least $10^4$ measurements have been performed.

The restriction of lattice sizes to $L\leq 18$ is mainly due to the costs
$\sim L^6$ in the determination of the monopole observable~(\ref{monopole})
because of the two summations over the space volume, one in the exponent, and
one in the Fourier-transformation of~(\ref{even_part},\ref{odd_part}).  The
statistical errors of the monopole masses are determined by the jack-knife
method using 16 blocks.

\subsection{Determination of the monopole mass}

\begin{figure}[t]
\begin{center}
  \psfig{file=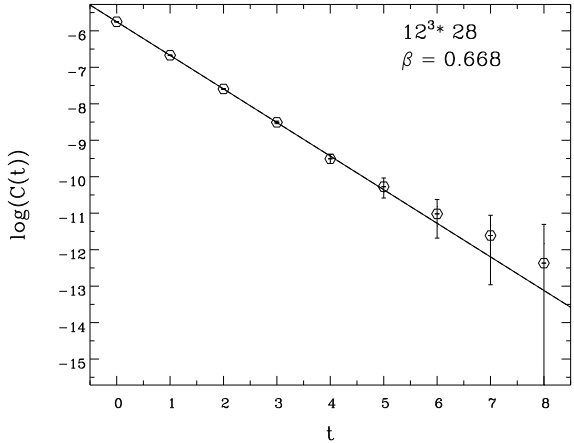,angle=90,width=\hsize}
  \caption{
    \label{corrlog}Logarithmic plot of the correlation function $C_+$ on
    the $L=12$ lattice at $\beta=0.668$. The solid curve is a fit
    to~(\ref{corrfunc}) with $m_+= 0.921(5)$ and $\rho = 0.0003(15)$.}
\end{center}
\end{figure}

We use the observables $\Phi_+$ and $\Phi_-$ with the lowest possible momenta,
i.e. $\vec p=0$ for $\Phi_+$ and $\vec p=(\pi/L,\pi/L,\pi/L)$ for $\Phi_-$.
The values of the monopole mass are obtained from their correlation function
which is assumed to have approximately the form
\begin{eqnarray}
\label{corrfunc}
  C_\pm(t)=\bra{0}\Phi_\pm^\star(t)\Phi_\pm(0)\ket{0}
  ={|\bra{0}\Phi_\pm(0)\ket{0}|}^2\cr
  +{|\bra{1}\Phi_\pm(0)\ket{0}|}^2\,
    \Bigl(e^{-E_1^\pm t}+e^{-E_1^\pm(T-t)}\Bigr).
\end{eqnarray}
The first term, the monopole condensate
\begin{equation}
    \label{rho}
    \rho=|\bra{0}\Phi_+\ket{0}|,
\end{equation}
is expected to vanish in the Coulomb phase in the infinite volume limit, but
should be allowed on finite lattices.

\begin{figure}[t]
\begin{center}
  \psfig{file=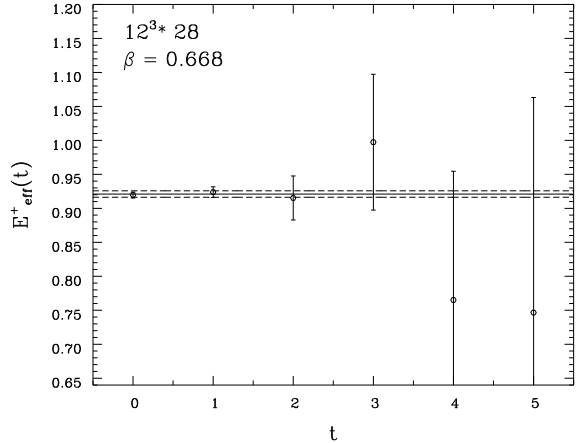,angle=90,width=\hsize}
  \caption{
    \label{correff}Effective energies corresponding to the correlation
    function from figure~\ref{corrlog}. The horizontal lines show the fit
    results for $E_1^+$ from~(\ref{corrfunc}) with $\rho=0$.}
\end{center}
\end{figure}

Figure~\ref{corrlog} shows as an example the data obtained on the $L=12$
lattice at $\beta=0.668$ with the fit by means of the correlation function
(\ref{corrfunc}).  The data in the whole range of $t$ are consistent with a
contribution of only one state to the correlation function. To verify this
more accurately, we have determined the effective energies $E_{\rm eff}$ with
$\rho$ both free and set equal to zero. The effective energies corresponding
to the correlation function from figure~\ref{corrlog} are plotted in
figure~\ref{correff}. They are stable with respect to $t$.  Similar results
are obtained for both operators (\ref{even_part}) and (\ref{even_part}) at all
investigated points in the Coulomb phase for all lattice sizes we have used.

Thus we find that within our numerical accuracy the $t$-dependence of the
correlation functions (\ref{corrfunc}) can be well described by a one particle
contribution. This is remarkable, as the Coulomb phase contains massless
photon which could in principle substantially complicate the form of the
correlation function (infra particle). It justifies the interpretation of the
observed energy as the energy or mass of the magnetic monopole. Furthermore,
there is no significant deviation from $\rho = 0$ and therefore from now on we
quote only results obtained under the assumption of vanishing condensate.

\subsection{Scaling behaviour of the monopole mass}

\begin{figure}[t]
\begin{center}
  \psfig{file=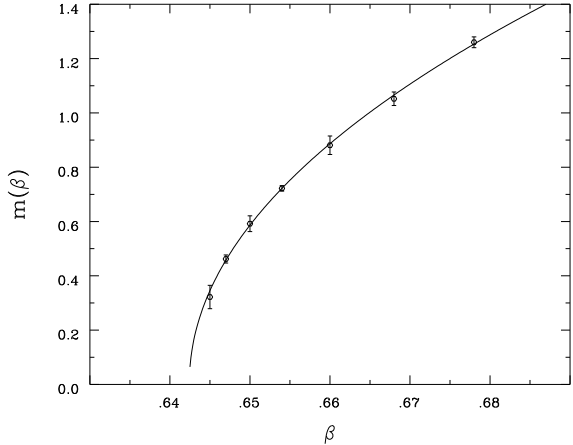,angle=90,width=\hsize}
  \caption{
    \label{monoscale} Scaling of the extrapolated monopole mass $m_\infty$
    with the power law~(\ref{m_infty}).}
\end{center}
\end{figure}

From the finite volume results for the masses of the $\Phi_+$ and $\Phi_-$
observables, we estimate the infinite volume mass $m_\infty$ as described
below in Sec.~\ref{volume}. The resulting values of $m_\infty$ are given in
Tab.~\ref{minfty} and Fig.~\ref{monoscale}.

\begin{table}[ht]
\begin{center}
\begin{tabular}{lr}
$\beta$ & $m_\infty$ \\
\hline
0.645   & 0.32(4) \\
0.647   & 0.46(2) \\
0.65    & 0.59(3) \\
0.654   & 0.72(2) \\
0.66    & 0.88(3) \\
0.668   & 1.05(3) \\
0.678   & 1.26(2) \\
\end{tabular}
\caption{\label{minfty} Extrapolation $m_\infty$ of the monopole
  mass to the infinite volume with~(\ref{second_ansatz}). The errors are
  estimated as described in the text.}
\end{center}
\end{table}

Figure~\ref{monoscale} demonstrates that the monopole mass $m_\infty$
decreases with decreasing $\beta$ and possibly vanishes at the phase
transition. The smallest value of $m_\infty$ which we obtained is about
$m_\infty\simeq 1/3$ (largest value of the corresponding correlation length is
$\xi_{\rm mon} \simeq 3$). In the range $m_\infty = 0.3 - 1.2$ a scaling
behaviour is observed, described by the power law (\ref{m_infty}) with the
critical exponent (\ref{nu_m}) and the critical point located at
\begin{equation}
 \label{b_c^Coul}  
    \beta_c^{\rm Coul}=0.6424(9).
\end{equation}

The value $\beta_c^{\rm Coul}$ is consistent with the result for $\beta_c$
obtained for the Villain action in~\cite{JeNe85,CoNe99}. The upper script
``Coul'' in (\ref{b_c^Coul}) indicates that the position of the critical point
has been obtained by means of an extrapolation of some observable from the
Coulomb phase. The value of the critical exponent $\nu\simeq 0.5$ has been
found earlier to describe the scaling behavior of the mass of a scalar gauge
ball in the pure U(1) lattice gauge theory with extended Wilson
action~\cite{CoFr98a,CoJe99a}.

It is important to realize why the smallest value of $m_\infty$ we obtained is
restricted.  As explained in the next section, an extrapolation of the finite
volume results for the monopole mass to the thermodynamic limit gets gradually
more and more difficult as the phase transition is approached. In fact, if the
finite volume dependence of the monopole masses were better understood, the
phase transition could be further approached without entering the region where
metastability occurs. On the $28^4$ lattice metastability occurs only at
$\beta\lesssim 0.6439$. An extrapolation of our data using (\ref{m_infty}) to
this $\beta$ suggests that this lattice size would allow to reach monopole
masses at least as small as 0.2. Thus the range of monopole mass values
investigated in this paper is restricted from below by an insufficient
understanding of finite size effects and not by the occurence of the two-state
signal.

\section{Volume dependence of the monopole mass}
\label{volume}

\begin{figure}[t]
\begin{center}
  \psfig{file=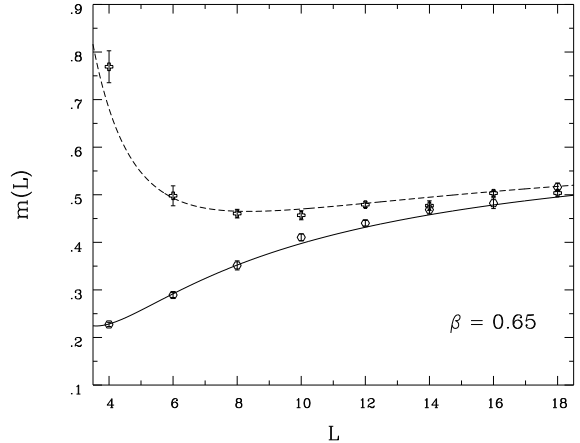,angle=90,width=\hsize}
  \caption{
    \label{monopole_mass_size} Dependence of the monopole masses $m_\pm$ on
    the finite spacial lattice size $L$ at $\beta=0.65$. The solid and dashed
    curves for $m_+$ and $m_-$, respectively, correspond to the
    ansatz~(\ref{first_ansatz}) up to the $1/L^2$ term.}
\end{center}
\end{figure}

Since in the Coulomb phase there exists a long range interaction mediated by
the massless photons, a strong dependence of the monopole masses on the finite
spacial size $L$ of the lattice is expected. Figure~\ref{monopole_mass_size}
displays the dependence of the masses $m_\pm$ obtained from the correlation
functions $C_+(t)$ (solid) and $C_-(t)$ (dashed) on the lattices of size
$L$.

The mass $m_+$ which is obtained from the symmetric combination of monopole
and antimonopole is smaller than $m_-$. The splitting of masses apparently
vanishes if the lattice size $L$ gets large, as expected from the degeneration
of the momenta of $\Phi_+$ and $\Phi_-$ in the infinite volume. However, for
smaller $L$ the functions $m_\pm(L)$ are rather complicated, $m_-(L)$ being
even nonmonotonic, and currently we do not have a sufficient theoretical
understanding of this $L$-dependence. To extrapolate to $L = \infty$ we
therefore combine the expected asymptotic behaviour of $m_\pm(L)$ with various
phenomenological ans\"atze.

It seems plausible, that at larger $L$, where both masses have already similar
values, the long-range Coulomb field of the monopole might dominate the
$L$-dependence. So we try to describe $m_\pm(L)$ at large $L$ by the classical
energy of a (magnetically) charged particle in a finite volume.

The classical energy of a (magnetic) charge $g$ with a Coulomb potential
$\phi(\vec r)=g/4\pi r$ in a spherical volume of radius $L$ in continuum is
\begin{equation}
\label{classical_sphere}
W_{\rm cont}(L)=\frac{\alpha_{\rm
    mag}}{2}\,\Bigl(\frac{1}{r_0}-\frac{1}{L}\Bigr),\qquad \alpha_{\rm
  mag}=\frac{g^2}{4\pi}.
\end{equation}
The diverging classical energy of the point particle has been regularized by a
restriction to $r\geq r_0$. The classical consideration gives a
characteristic $1/L$ dependence 
\begin{equation}
      \label{clas_mL}
 m_\pm(L)=m_\infty-\frac{c_1}{L}.
\end{equation}
with some value of the coefficient $c_1$ depending on the shape of the finite
volume, boundary conditions and regularization. There is no difference between
$m_+$ and $m_-$ at the classical level. The splitting of masses is probably a
quantum mechanical effect, which we have not estimated.

In order to obtain the value of $c_1$ for the cubic lattices used in the
simulations, we first carry out a classical computation analogous to
(\ref{classical_sphere}) in the lattice regularization. We use the
lattice Coulomb potential
\begin{eqnarray}
  \phi_{\rm latt}(\vec r)=\frac{g}{L^3}\sum_{\vec p\neq 0}
    \frac{\exp(i\vec p\cdot \vec r)}{2\sum_{\mu=1}^3(1-\cos p_j)},\cr
    p_j=\frac{2\pi}{L}n_j,\quad n_j\in\{0,\ldots,L-1\}
\end{eqnarray}
on a $L^3$ cubic lattice with periodic boundary conditions. In the computation
of the energy, the gradient is replaced by a finite difference, and the
integral over the volume is replaced by a finite sum. We determine the energy
numerically. The result is very well approximated by
\begin{equation}
  \label{w_latt}
     W_{\rm latt}(L)=0.2524-1.3881\,\frac{\alpha_{\rm mag}}{2\,L}.
\end{equation}
Compared with~(\ref{classical_sphere}), only the pre-factors have been
changed. We do not specify any errors of the coefficients because the
statistical errors from the simulations are larger by an order of magnitute.
The resulting estimate of the $L$-dependence of the monopole mass on a
finite $L^3$ lattice is therefore (\ref{clas_mL}) with
\begin{equation}
      \label{classcoeff}
  c_1=1.3881\,\frac{\alpha_{\rm mag}}{2}.
\end{equation}
The drawback of this estimate is the fact that the antiperiodic boundary
conditions on the gauge field are not fully respected.

Another way to estimate the value of $c_1$ is to interpret the antiperiodic
boundary conditions for the gauge field as a three-dimensional infinite cubic
lattice of alternating charges, the distance between nearest neighbours being
$L$. For the continuum space this consideration leads to the problem of
lattice sums, relevant for various crystaline materials. Taking over the well
known results from the condensed matter physics \cite{Ki71} implies that the
energy of one monopole in the field of all others is equal to the second term
in (\ref{clas_mL}) with
\begin{equation}
      \label{madcoeff}
  c_1=1.7476\,\frac{\alpha_{\rm mag}}{2}.
\end{equation}
The numerical factor in (\ref{madcoeff}) is the Madelung constant. It
determines the classical energy of cubic ion crystals, though its actual
calculation requires particular mathematical attention \cite{Ki71}. We prefer
(\ref{madcoeff}) to (\ref{classcoeff}) because it respects the correct
boundary conditions for the gauge field.

An estimate of $\alpha_{\rm mag}$ can be obtained from the numerical analysis
of the static potential of U(1) lattice gauge theory with Villain action by
using the Dirac relation
\begin{equation}
\label{dirac}
  g\cdot e=2\pi\qquad\Rightarrow\qquad 
  \alpha_{\rm mag}=\frac{1}{4\,\alpha_{\rm el}}.
\end{equation}

\begin{table}[ht]
\begin{center}
\begin{tabular}{llll}
$\beta$ & $\alpha_{\rm el}$ & $\alpha_{\rm mag}$ & $c_1$ \\
\hline
0.645 & 0.1836 & 1.3614 & 1.1896 \\
0.647 & 0.1766 & 1.4159 & 1.2372 \\
0.650 & 0.1687 & 1.4817 & 1.2947 \\
0.654 & 0.1606 & 1.5571 & 1.3606 \\
0.660 & 0.1508 & 1.6580 & 1.4487 \\
0.668 & 0.1402 & 1.7829 & 1.5579 \\
0.678 & 0.1293 & 1.9329 & 1.6889 \\
\end{tabular}
\end{center}
\caption{
  \label{alphas} Dual renormalized coupling $\alpha_{\rm mag}$ obtained
  from~(\ref{dirac}). $c_1$ is the coefficient~(\ref{madcoeff}) used in the 
  ansatz~(\ref{second_ansatz}).}
\end{table}

We use the values of the renormalized electrical coupling $\alpha_{\rm
  el}=e^2/4\pi$ from~\cite{JeNe85}. Table~\ref{alphas} contains the values of
$\alpha_{\rm el}$ and $\alpha_{\rm mag}$ for different $\beta$.

The fact that classical energy considerations suggest a $1/L$ dependence
motivates our ans\"atze in terms of a polynomial in $1/L$. The necessary scale
is assumed to be provided by the infinite volume mass $m_\infty$ itself. Our
first ansatz is
\begin{equation}
\label{first_ansatz}
  m_\pm(L)=m_\infty+\frac{c_1^{(\pm)}}{L}+\frac{c_2^{(\pm)}}{m_\infty L^2}
  +\frac{c_3^{(\pm)}}{m_\infty^2 L^3}\cdots
\end{equation}
To test the estimate (\ref{w_latt}), different coefficients $c_i^{(+)}$ and
$c_i^{(-)}$ are allowed even for $i=1$. Because of the theoretical
uncertainty in this and the other ans\"atze, we have to be very cautious in
estimating the errors of $m_\infty$.

Fig.~\ref{monopole_mass_size} shows the fits (\ref{first_ansatz}) upto
$1/L^2$. The results obtained for $m_\infty$ become stable if the fit is
restricted to data with $L\geq 8$. In this case, $m_\infty$ is not
signifficantly changed if more points at small $L$ are omitted or if $1/L^3$
contributions are included. The coefficients $c_1^{(+)}$ and $c_1^{(-)}$
roughly agree, but differ from the classical value $c_1$ obtained from
(\ref{madcoeff}) and given in
table~\ref{alphas} by a factor of $1.2 - 1.6$. 

\begin{figure}[t]
\begin{center}
  \psfig{file=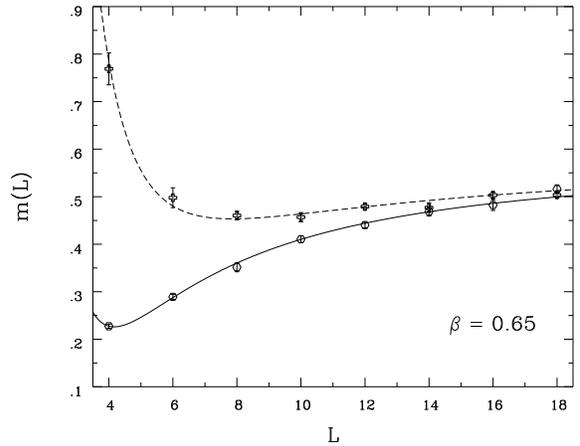,angle=90,width=\hsize}
  \caption{
    \label{monopole_mass_size2} Dependence of the masses of $m_+$ (solid
    line) and $m_-$ (dashed line) on the lattice size $L$ at $\beta=0.65$. The
    curves correspont to~(\ref{second_ansatz}) up to the term $1/L^3$.}
\end{center}
\end{figure}

Nevertheless, it is possible to fix $c_1^{(+)}=c_1^{(-)}=c_1$ to the classical
value and try another ansatz
\begin{equation}
\label{second_ansatz}
  m_\pm(L)=m_\infty-\frac{c_1}{L}+\frac{c_2^{(\pm)}}{m_\infty L^2}
  +\frac{c_3^{(\pm)}}{m_\infty^2 L^3}\cdots.
\end{equation}
We have used this ansatz with both estimates of the coefficient $c_1$, given
in (\ref{classcoeff}) and (\ref{madcoeff}). It turned out that the results for
the monopole mass in the infinite volume are consistent within the error bars.
In the following we thus describe only the results obtained using the Madelung
constant in (\ref{madcoeff}).

Fig.~\ref{monopole_mass_size2} shows the same data as
fig.~\ref{monopole_mass_size}, but with the functions~(\ref{second_ansatz}) up
to $1/L^3$. The fit results $m_\infty$ become stable if the fit includes the
$1/L^3$ contributions. A restriction to large $L$ is necessary only
for $\beta$ close to the phase transition.

We consider the values of $m_\infty$ obtained in the fit by means of
(\ref{second_ansatz}) with all listed terms as the best determination of the
monopole mass in the infinite volume and list the results in
table~\ref{minfty}.  To obtain an estimate of the errors of $m_\infty$
obtained in this way, we compare the fit results from~(\ref{second_ansatz})
with and without higher $1/L^k$ terms. Further we omit various number of
points at small $L$ from the fits. The error of $m_\infty$ is then estimated
from the variation of the fit results under the different conditions. The
values of these errors are also given in table~\ref{minfty}. Using
instead of (\ref{second_ansatz}) the ansatz (\ref{first_ansatz}) results in
the values of $m_\infty$ compatible with those in table~\ref{minfty} within the
listed errors.

\begin{figure}[t]
\begin{center}
  \psfig{file=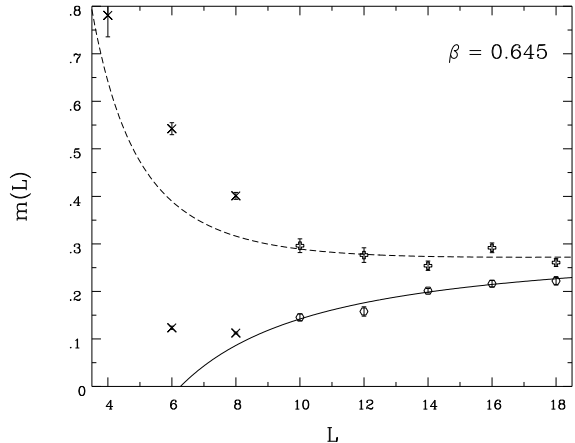,angle=90,width=\hsize}
  \caption{
    \label{worstcase} Dependence of the masses of $m_+$ (solid line) and
    $m_-$ (dashed line) on the lattice size $L$ at $\beta=0.645$ which is
    the point closest to the phase transition. The curves correspond
    to~(\ref{second_ansatz}) up to $1/L^2$. Data points with $L<10$ have been
    excluded from the fits.}
\end{center}
\end{figure}

Though insufficiently motivated, our extrapolation procedure is rather stable
at the $\beta$ values farther from the phase transition. Even the use of the
coefficient $c_1$ given in (\ref{classcoeff}) instead of (\ref{madcoeff}),
does not change the results significantly. The extrapolation is most
problematic at the point closest to the phase transition, $\beta=0.645$.
Figure~\ref{worstcase} shows the $L$-dependence of the $m_\pm$ masses in this
case, using (\ref{madcoeff}).  Since the finite volume mass splitting between
$m_+$ and $m_-$ remains significantly nonzero on the largest lattice, the
systematic error is probably quite large. From the different fits we estimate
it to be about $20\%$ at this $\beta$. In order to reduce this error and to
further approach the phase transition, larger lattices and a more reliable
extrapolation procedure would be needed.

\section{Results in the confinement phase}

\subsection{Simulations and statistics}

Our simulations in the confinement phase are performed with the same algorithm
and the same boundary conditions as described in Sec.~\ref{simstat}. The
lattice volumes are $L^3\cdot T$ with $T=24$ fixed and several $L$. At each
$\beta$, at least 5000 measurements have been made starting with an ordered
system and 5000 starting with a completely disordered system. Before the
measurements we have used $2.5\cdot 10^3$ (away from the phase transition) to
$2.5\cdot10^4$ (close to the transition) sweeps for thermalization. The
monopole operators~(\ref{even_part}) and~(\ref{odd_part}) are measured after
each $25$ update sweeps.

\begin{table}[ht]
\begin{center}
\begin{tabular}{lrr}
$\beta$ & $L_{\rm max}$ & $\tau_{\rm int}$ \\
\hline
0.6     & 12 &  1.1 \\
0.62    & 14 &  1.2 \\
0.625   & 14 &  1.3 \\
0.635   & 14 &  1.3 \\
0.638   & 12 &  2.6 \\
0.64    & 16 &  2.9 \\
0.641   & 16 &  4.2 \\
0.642   & 16 &  5.9 \\
0.6425  & 18 &  7.2 \\
0.643   & 14 &   10 \\
0.6435  & 18 &   15 \\
0.6436  & 18 &   30 \\
0.6437  & 18 &   35 \\
\end{tabular}
\caption{
  \label{confstat} Values of $\beta$ at which the simulations in the confinement
  phase were performed. The lattice sizes are $L=4,6,\ldots,L_{\rm max}$. 
  $\tau_{\rm int}$ denotes the integrated autocorrelation time.}
\end{center}
\end{table}

Table~\ref{confstat} displays the parameters of the simulations and the
integrated autocorrelation times $\tau_{\rm int}$ of the action density in
multiples of $25$ sweeps. It has been determined on the $L=8$ lattice at
$\beta<0.6436$ and on $L=12$ at $0.6436\leq\beta$. The integrated
autocorrelation of the monopole condensate is compatible with these values.

The $\beta$ values are chosen far enough from the phase transition, so that
the flips between both phases at the phase transition do not occur in the
runs. On the largest lattices we could simulate this leads to the exclusion of
runs at $\beta=0.6438$.

Because of the broken $\mathbf{Z}_2$ symmetry, the finite system
sometimes flips between the two possible ground states within the confinement
phase. The flip probability increases with decreasing lattice volume and
when the phase transition is approached. We observe flips at
$\beta\leq 0.643$ only on the small lattices with $L\leq 8$. At
$\beta=0.6435$, we observe them up to the $L=14$ and at $\beta=0.6436,0.6437$
on all lattice sizes up to $L=18$. On our largest lattices, the system does
not flip more than one or two times during the whole run. In these cases we cut the
corresponding parts of the runs to ensure that the unwanted intermediate
states do not contribute to the resulting value of the monopole condensate.
However, runs with more frequent flips are discarded.

\subsection{Monopole condensate}

The monopole condensate $\bra{0}\Phi_+\ket{0}$ is measured directly, and its
modulus (\ref{rho}) can be determined from the correlation
function $C_+(t)$ of $\Phi_+$~(\ref{corrfunc}). Both methods yield compatible results.

\begin{figure}[t]
\begin{center}
  \psfig{file=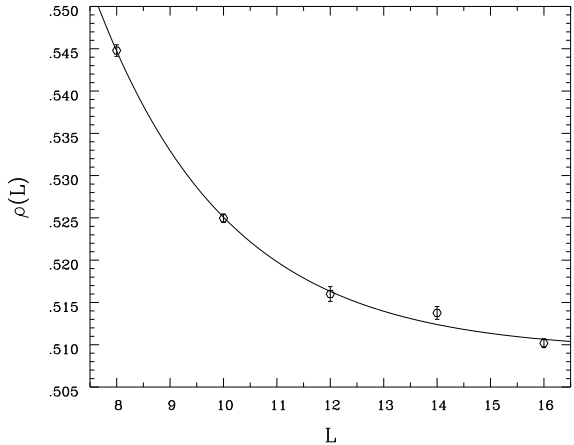,angle=90,width=\hsize}
  \caption{
    \label{cond1} Dependence of the monopole condensate on  $L$ at
    $\beta=0.64$.  The curve corresponds to~(\ref{condscale}). The scale of
    the $y$-axis is very fine.}
\end{center}
\end{figure}

The value of $\rho$ does not depend on the lattice extent $T$ in time
direction, but it is weakly dependent on its spacial extent $L$. The
$L$-dependence of $\rho$ can be described with an exponential law
\begin{equation}
\label{condvolume}
\rho(L)=\rho_\infty+a\,e^{-bL}
\end{equation}
with constants $a$, $b$ and the infinite volume value $\rho_\infty$.
Figure~\ref{cond1} displays $\rho(L)$ at $\beta=0.64$ with the
fit~(\ref{condvolume}) used for the extrapolation.

\begin{figure}[t]
\begin{center}
  \psfig{file=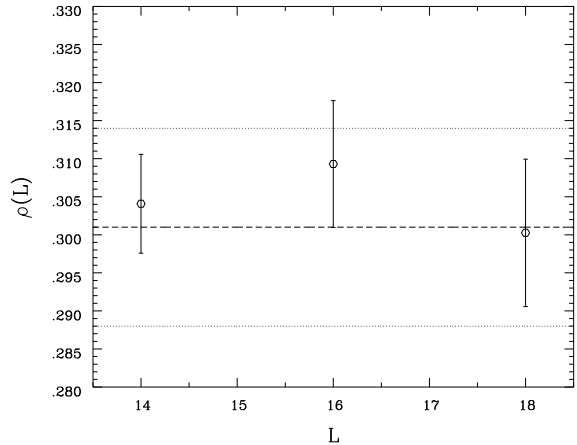,angle=90,width=\hsize}
  \caption{
    \label{cond2} Dependence of the monopole condensate on  $L$ at 
    $\beta=0.6436$ (close to the phase transition). The horizontal lines
    indicate the estimated value $\rho_\infty$ and its error.}
\end{center}
\end{figure}

At $\beta=0.6436$ and $\beta=0.6437$ the data for the condensate could
be obtained only on large lattices due to the ground state flips on smaller
ones. As there is no significant $L$-dependence of $\rho$ on the largest
lattices, we take the average of the obtained values to represent
$\rho_\infty$ instead of using the ansatz~(\ref{condvolume}).
Figure~\ref{cond2} shows $\rho(L)$ at $\beta=0.6436$ and the resulting value
of $\rho_\infty$.

\begin{table}[ht]
\begin{center}
\begin{tabular}{ll}
$\beta$ & $\rho_\infty$ \\
\hline
0.6     & 0.823(4)  \\
0.62    & 0.730(8)  \\
0.635   & 0.602(5)  \\
0.638   & 0.560(3)  \\
0.64    & 0.509(2)  \\
0.641   & 0.480(6)  \\
0.642   & 0.441(4)  \\
0.6425  & 0.414(5)  \\
0.643   & 0.383(3)  \\
0.6435  & 0.325(10) \\
0.6436  & 0.300(13) \\
0.6437  & 0.275(10) \\
\end{tabular}
\caption{\label{condtab} Extrapolation of the monopole condensate $\rho$ to
  the infinite spacial volume.} 
\end{center}
\end{table}

Table~\ref{condtab} lists the extrapolated monopole condensate values
$\rho_\infty$ at different values of $\beta$.

\subsection{Scaling of the monopole condensate}

\begin{figure}[t]
\begin{center}
  \psfig{file=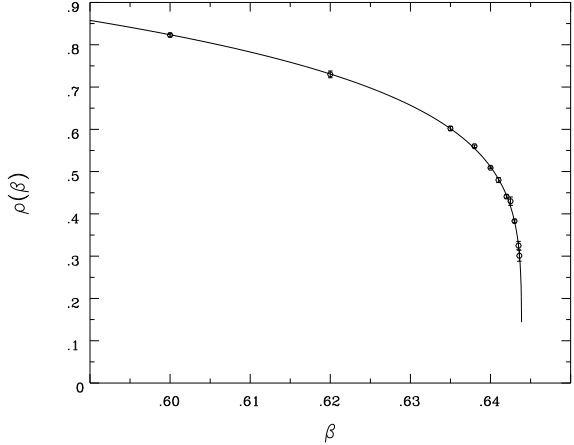,angle=90,width=\hsize}
  \caption{
    \label{condfig} Scaling behavior of the monopole condensate extrapolated 
    to $L = \infty$ with $\beta$. The curve corresponds to the power
    law~(\ref{condlaw}).}
\end{center}
\end{figure}

\begin{figure}[t]
\begin{center}
  \psfig{file=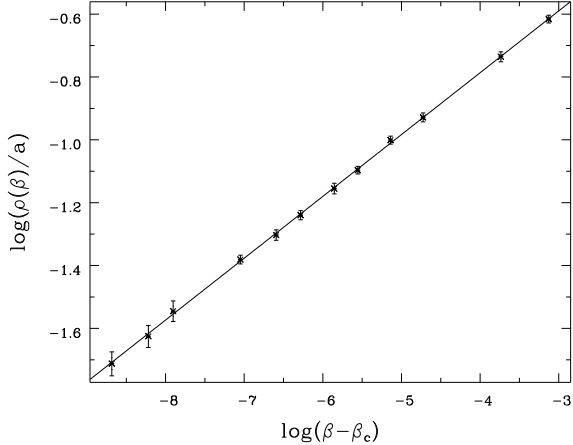,angle=90,width=\hsize}
  \caption{
    \label{condlog} Scaling of the monopole condensate extrapolated to 
    $L = \infty$ with $\beta$ in a logarithmic diagram. The line corresponds
    to the power law~(\ref{condlaw}).}
\end{center}
\end{figure}

The $\beta$-dependence of the monopole condensate $\rho_\infty$ is 
described with an astonishing precision by a simple power law
\begin{equation}
\label{condlaw}
\rho_\infty(\beta)=a_\rho{(\beta_c^{\rm conf}-\beta)}^{\beta_{\rm exp}}
\end{equation}
with the value (\ref{magneticexp}) of the magnetic exponent and
\begin{equation}
  \label{b_c^conf}
  \beta_c^{\rm conf} = 0.6438(1).
\end{equation}
Figure~\ref{condfig} shows the scaling of the monopole condensate with the
power law~(\ref{condlaw}). The logarithmic plot, Fig.~\ref{condlog}, is
obtained by plotting $\log(\rho(\beta)/a_\rho)$ versus
$\log(\beta-\beta_c^{\rm conf})$ with $a$ and $\beta_c^{\rm conf}$ taken as
the fit results from~(\ref{condlaw}).

We note that the value (\ref{b_c^conf}) of $\beta_c^{\rm conf}$ and the value
(\ref{b_c^Coul}) of $\beta_c^{\rm Coul}$ are consistent within two error bars. The value of
$\beta_c^{\rm conf}$ is much more precise, because the determination of the
condensate in the confinement phase is much more precise than that of the
monopole mass in the Coulomb phase.

\subsection{Excited states of the condensate}

From the correlation function~(\ref{corrfunc}) in the confinement phase, one
can determine in addition to the monopole condensate the energy and amplitude
of a first excited state. If these values gave reasonable results in the
infinite volume, they would correspond to a magnetically charged particle-like
excitation of the monopole condensate. 

There have been attempts to determine the properties of such a
state~\cite{DiLu98}. However, since we use different boundary conditions than
the authors of~\cite{DiLu98}, and the remaining symmetry that is dynamically
broken is $\mathbf{Z}_2$ in our case instead of the magnetic U(1), our results
are difficult to compare with those presented in~\cite{DiLu98}.

We examine the dependence of the energy $E_1^{(\pm)}$ and amplitude squares
$a_\pm={|\bra{1}\Phi_\pm\ket{0}|}^2$ on the finite spacial lattice size $L$.
The corresponding masses $m_\pm(L)$ obtained from $E_1^{(\pm)}$ by means of
the dispersion relations (\ref{m_+}) and (\ref{dispersion}) show no clear $L$
dependence that would allow an extrapolation to the infinite volume. Also the
masses $m_+$ and $m_-$ do not approach each other as $L$ is increased.

\begin{table}[ht]
\begin{center}
\begin{tabular}{llllll}
$m_+(L)$&\multicolumn{5}{c}{$L$}\\
\hline
$\beta$ & 10       & 12       & 14       & 16      & 18       \\ 
\hline
0.6     & 1.90(3)  & 1.95(2)  &          &         &          \\
0.62    & 1.43(2)  & 1.49(2)  & 1.53(2)  &         &          \\
0.635   & 0.93(1)  & 0.94(1)  & 0.98(2)  &         &          \\
0.638   & 0.77(1)  & 0.78(1)  &          &         &          \\
0.64    & 0.63(2)  & 0.635(8) & 0.67(1)  &         &          \\
0.641   & 0.554(8) & 0.565(7) & 0.582(7) & 0.61(1) &          \\
0.642   & 0.461(7) & 0.479(8) &          & 0.50(1) &          \\
0.6425  & 0.396(8) & 0.439(5) & 0.440(7) & 0.44(1) & 0.446(3) \\
0.643   & 0.31(1)  & 0.29(2)  & 0.34(1)  &         &          \\
0.6435  & 0.21(2)  & 0.20(3)  & 0.18(2)  & 0.25(3) & 0.26(2)  \\
0.6436  &          &          & 0.16(2)  & 0.20(2) & 0.19(2)  \\
0.6437  &          &          & 0.16(2)  & 0.14(2) & 0.16(2)  \\

\end{tabular}
\caption{\label{excited1} Mass $m_+$ of the even excited state in the
  confinement phase.}
\end{center}
\end{table}

\begin{table}[ht]
\begin{center}
\begin{tabular}{llllll}
$m_-(L)$&\multicolumn{5}{c}{$L$}\\
\hline
$\beta$ & 10       & 12       & 14       & 16       & 18       \\ 
\hline
0.6     & 2.11(2)  & 2.15(2)  &          &          &          \\
0.62    & 1.70(1)  & 1.68(2)  & 1.69(2)  &          &          \\
0.635   & 1.23(1)  & 12.3(1)  & 1.29(1)  &          &          \\
0.638   & 1.08(1)  & 1.09(1)  &          &          &          \\
0.64    & 0.99(1)  & 0.98(1)  & 0.96(1)  &          &          \\
0.641   & 0.906(7) & 0.885(6) & 0.890(4) & 0.864(5) &          \\
0.642   & 0.810(7) & 0.801(9) &          & 0.781(9) &          \\
0.6425  & 0.764(6) & 0.762(4) & 0.731(8) & 0.716(6) & 0.725(3) \\
0.643   & 0.719(9) & 0.66(1)  & 0.66(1)  &          &          \\
0.6435  & 0.58(1)  & 0.60(2)  & 0.51(2)  & 0.52(1)  & 0.48(3)  \\
0.6436  &          &          & 0.50(1)  & 0.50(2)  & 0.48(2)  \\
0.6437  &          &          & 0.46(2)  & 0.45(2)  & 0.42(4)  \\
\end{tabular}
\caption{\label{excited2} Mass $m_-$ of the odd excited state in the
  confinement phase.}
\end{center}
\end{table}

Tables~\ref{excited1} and~\ref{excited2} show the masses $m_+$ and $m_-$,
respectively, for various $\beta$ and the spacial lattice sizes $L$.

\begin{figure}[t]
\begin{center}
  \psfig{file=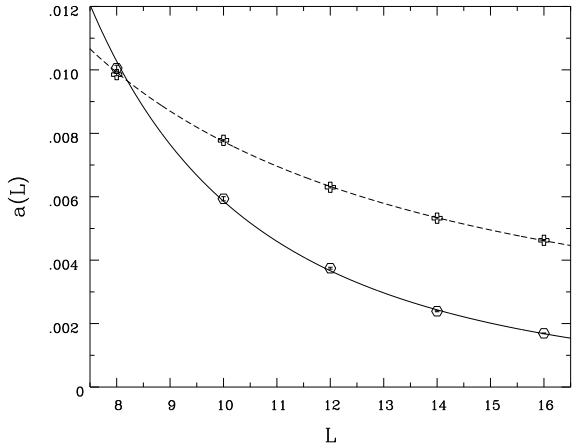,angle=90,width=\hsize}
  \caption{
    \label{excitedfig}Amplitudes $a_\pm(L)$ of the $\Phi_+$ (solid) and
    $\Phi_-$ (dashed) excitations depending on $L$ with the
    curves~(\ref{extrapol}).} 
\end{center}
\end{figure}

The amplitudes $a_\pm(L)$ decrease as the spacial lattice size is
increased. We find that a power law
\begin{equation}
\label{extrapol}
  a_\pm(L)=a_\infty^{(\pm)}+\frac{c_\pm}{L^r_\pm}
\end{equation}
gives values of the exponent $r$ in the range $1\ldots3$ and values
$a_\infty^{(\pm)}$ consistent with zero. Fig.~\ref{excitedfig} shows as an
example the amplitudes $a_\pm(L)$ at $\beta=0.64$ with fits by the
functions~(\ref{extrapol}).

These results indicate, that the observed excitations are not present in the
infinite volume limit and the corresponding quantities $m_\pm$ should not be
interpreted physically as particle masses. For this reason the question of an
appropriate renormalization of the monopole condensate remains unanswered.

\section{Conclusion and discussion}

In the pure U(1) gauge theory with the Villain action we have investigated the
scaling behaviour of the monopole mass in the Coulomb phase and of the
monopole condensate in the confinement phase. Both observables indicate a
critical behavior in the vicinity of the phase transition between these
phases, with the values of the corresponding exponents (\ref{nu_m}) and
(\ref{magneticexp}). Assuming that a continuum theory can be constructed at
this phase transition, these results indicate that the monopoles appear also
in such a theory. In particular, the monopole mass in the Coulomb phase can
vanish or stay finite nonzero in physical units. The Gaussian value
(\ref{nu_m}) of the exponent suggests (though not implies) that the
corresponding continuum theory may be trivial. This property would then hold
also for the scalar QED in the Coulomb phase, as this theory is obtained by
duality transformations from the pure compact U(1) theory \cite{Pe78,FrMa}.

As for the question of the existence of a continuum limit, in this work we do
not contribute to the resolution of the controversy whether the phase
transition is of the second or of the weak first order \cite{ArLi99}. This was
also not our aim in this paper, because for this purpose different methods
would be more appropriate. As the values (\ref{b_c^Coul}) and (\ref{b_c^conf})
of $\beta_c$ obtained from the Coulomb and the confinement phases are
consistent within the error bars times 1.5, our results are consistent with
the second order.  However, a small difference $|\beta_c^{\rm conf} -
\beta_c^{\rm Coul}| \approx 0.001$ is not excluded by our data, allowing a
weak first order transition for the Villain action.

What we want to point out is that, in spite of such a possibility, which
actually never can be excluded with full certainty, the pure compact QED on
the lattice remains to be a candidate for the construction of an
interesting quantum field theory in continuum. The scaling behaviour of the
monopole mass and in particular of the monopole condensate is of a high
quality, described by a single exponent in the whole accessible region of
values of the observables. The same holds also for various other observables
\cite{JeLa,CoFr97b,CoJe99a,CoJe99b}.  This suggests the existence of a
continuous phase transition somewhere in the parameter space of possible
lattice versions of the pure compact QED. This point of view might be relevant
also for the compact lattice QED with fermions \cite{CoFr98b}. Therefore the
compact QED merits further investigation with larger resources and improved
methods.

\vspace{1cm}{\bf ACKNOWLEDGEMENTS} \\

We thank U.-J. Wiese for numerous discussions, suggestions and for providing
us with his program which we partly used. Discussions with J. Cox and A. Di
Giacomo are acknowledged. J.J. and H.P. thank NIC J\"ulich (former HLRZ
J\"ulich), where the computations have been performed, for hospitality. T.N.
thanks the Helsinki Institute of Physics for hospitality.

\bibliographystyle{wunsnot}   


\end{document}